\documentclass{article}
\usepackage[utf8]{inputenc}
\usepackage{authblk}
\usepackage{setspace}
\usepackage[margin=1.25in]{geometry}
\usepackage{graphicx}
\graphicspath{ {./figures/} }
\usepackage{subcaption}
\usepackage{amsmath}
\usepackage{lineno}
\usepackage{multirow}
\usepackage{multicol}
\usepackage{algpseudocode}
\usepackage{algorithm}
\usepackage{booktabs}
\usepackage{makecell}
\usepackage{threeparttable}
\usepackage{rotating}
\usepackage{lineno}

\usepackage[table]{xcolor}

\usepackage[style=ieee,=
citestyle=numeric-comp,
sorting=none]{biblatex}
\title{Universal Neural Network Based Calibration and Control of Programmable Classical and Quantum Photonic Integrated Processors}

\author[1*]{José Roberto Rausell-Campo}
\author[2]{Daniele Melati}
\author[3]{Bhavin Shastri}
\author[4]{Daniel Pérez-López}
\author[1]{José Capmany Francoy}

\affil[1]{Photonics Research Lab, iTEAM, Universitat Politècnica de València, Camí de Vera S/N, 46022 Valencia, Spain}
\affil[2]{Centre de Nanosciences et de Nanotechnologies, CNRS, Université Paris-Saclay, 91120 Palaiseau, France}
\affil[3]{Centre for Nanophotonics, Department of Physics, Engineering Physics and Astronomy, Queen’s University, Kingston, ON K7L 3N6, Canada}
\affil[4]{IPronics Programmable Photonics S.L, Av. de Blasco Ibáñez, 25, Valencia, Spain}
\affil[*]{Corresponding author. Email: joraucam@upv.es}

\date{}

\begin{document}
\maketitle

\begin{abstract}
Efficient calibration and control of programmable photonic integrated circuits are fundamental for scaling quantum and classical optical computing processors. While neural network-based models offer an architecture-agnostic solution, existing approaches suffer from limited learning and generalization capabilities due to the many-to-one mapping problem between sets of control signals and optical responses, and biased training datasets derived from uniform current sampling. In this work, we propose a universal calibration and control framework employing tandem neural networks combined with two novel data generation strategies: architecture-aware sampling based on Haar measure principles, and optimized sampling, a physics-agnostic approach utilizing differential evolution. We experimentally validate these methods on $3\times3$ and $4\times4$ coherent MZI meshes, demonstrating that our approach addresses the sampling bias inherent in previous works. When evaluated using random unitary matrices, our solution outperforms standard uniform sampling baselines by $\sim 2$ bits of precision. Furthermore, we experimentally extend the application of this framework to coherent detection, achieving precise control over both amplitude and phase, and validate its impact on photonic neural network tasks. 
\end{abstract}
\newpage

\newpage
\section{Introduction}
Programmable photonic integrated circuits (PICs) \cite{Bogaerts2020} are on-chip optical systems that precisely manipulate light and whose response can be dynamically controlled. This reconfigurability is enabled by tunable building blocks, most commonly Mach-Zehnder interferometers (MZIs), which steer light via interference between distinct optical paths. Programmable PICs can implement linear transformations \cite{doi:10.1126/science.aab3642, Harris:18}, which are fundamental to quantum computing \cite{Harris2017}, artificial intelligence \cite{Bandyopadhyay2024Single-chipTraining, Shen2017}, combinatorial optimization \cite{Prabhu2020}, matrix inversion \cite{Prabhu2020, Cavicchioli:24}, optical mode processing \cite{Annoni2017, Milanizadeh2022SeparatingProcessor, SeyedinNavadeh2024}, microwave photonics \cite{10092945, Perez-Lopez2024}, and analog photonic computing \cite{https://doi.org/10.1002/lpor.202200360}. Photonic unitary linear transformations are implemented by cascading multiple MZIs, each with phase shifters, arranged in a planar configuration. Typically, these MZIs are arranged in triangular or rectangular layouts \cite{Clements2016, Reck1994}. A schematic of the rectangular architecture, along with its basic building block, is illustrated in Fig.~\ref{fig:conceptual}\textbf{a}.

Scaling these systems for real-world applications requires precise calibration and control of their constituent building blocks. This is essential because, as system dimensionality increases, mismatches in local components propagate through the network, degrading global performance. Calibration establishes the relationship between the voltages or currents applied to the phase actuators and the resulting optical response \cite{Alexiev2021}. Accurate calibration requires sequential characterization of each MZI in the circuit, including measurement of individual transfer functions while accounting for the parasitic effects and crosstalk from adjacent components. Control consists of determining the set of phases required to implement a given unitary matrix. However, standard decomposition algorithms for linear processors assume ideal components and do not account for fabrication-induced deviations \cite{Clements2016}, leading to imbalances in the beam-splitter ratios of the MZIs.  These errors, combined with other undesired effects like thermal crosstalk \cite{8610179}, produce system responses that differ from the intended target operation. This difference has been shown to increase exponentially with circuit size, significantly reducing accuracy in tasks such as optical neural networks and quantum computing \cite{Hamerly2022AsymptoticallyPhotonics, PhysRevA.92.032322}.  

Here, we present a universal, black-box neural network framework and experimentally demonstrate the calibration and control of programmable photonic meshes. We show that conventional methods based on uniform current sampling fail to generalize effectively. To address this, we introduce improved sampling strategies leveraging Haar initialization, a uniformly random unitary transformation, and differential evolution, a gradient-free optimization algorithm that iteratively mutates and selects candidate solutions. 
Our methodology provides a scalable, architecture-independent framework toward robust photonic processors.

In literature, hardware error-correction techniques incorporate the nonideal behavior of Mach-Zehnder interferometer beam splitters directly into the control stage \cite{Bandyopadhyay2021, Kumar2021MitigatingLO}. However, individually characterizing these errors introduces complexity into the calibration process. Alternative architectural approaches introduce redundant tunable MZIs within the basic building blocks to mitigate these imperfections, increasing the number of control signals, insertion losses, and susceptibility to thermal crosstalk \cite{Hamerly2022AsymptoticallyPhotonics, Wang:20, Miller:15}.

To avoid explicit error characterization, self-configuring algorithms \cite{Miller:13, 9975587, Perez-Lopez2020, PhysRevApplied.22.054011, Dhand_2016} and data-driven machine learning techniques have emerged as alternative routes for high-fidelity linear transformations. While self-configuring methods inherently compensate for hardware degradation by iteratively minimizing a loss function, they are computationally demanding and lack the agility required for fast matrix updates. Similarly, data-driven strategies, ranging from clear-box virtual replicas requiring precise error modeling \cite{Fyrillas:24} to grey-box neural networks that characterize the chip via architectural priors \cite{Youssry2024ExperimentalControl} reduce some calibration bottlenecks. Nevertheless, achieving precise chip control still requires coupling the neural network with a secondary optimization stage, inheriting the scalability and speed limitations of self-configuring algorithms.

Alternatively, architecture-agnostic black-box models can capture hardware nonidealities without prior physical knowledge. While early demonstrations using feedforward neural networks (FNNs) achieved precise PIC calibration \cite{Cem:23, PhysRevApplied.15.044003}, they still require separate decomposition algorithms for chip control. Direct FNN-based control has been proposed \cite{10.1063/5.0293082}, but this approach suffers from the many-to-one mapping problem, where multiple phase configurations yield identical optical responses (Fig.~\ref{fig:conceptual}\textbf{c}), leading to poor training convergence and severe data-dependency. To resolve this ambiguity, tandem neural network architectures have been introduced \cite{10416579, Fan:25}. By coupling a forward network, which models the hardware response,  with an inverse network, which predicts the required control phases, this configuration acts as a regularizer as shown in Fig.~\ref{fig:conceptual}\textbf{b}. It ensures the learning of unique, mathematically consistent solutions, enabling efficient, calibration-aware control of MZI meshes.

However, these studies share a significant limitation: the training dataset used for both forward and inverse networks are generated by uniformly sampling the currents applied to the phase actuators. In coherent meshes, this sampling method produces skewed distributions of the resulting unitary matrices \cite{PhysRevApplied.11.064044}. This issue is not apparent in prior simulation or experimental results because the trained models are evaluated on test sets drawn from the same distribution as the training set. A universal model, however, must predict accurate control settings for all matrices in the unitary space, which typically follow a distribution different from that produced by uniform current sampling. Consequently, when evaluated the models on random matrices outside the training distribution, model accuracy decreases substantially relative to results on the original test sets.

Here, we use tandem neural networks to simultaneously calibrate and control coherent programmable MZI meshes. We experimentally demonstrate that prior methods based on uniform sampling of currents are insufficient for training a generalizable models. To overcome this limitation, we introduce two sampling strategies: an architecture-aware data-generation based on Haar-initialization principles, and an optimized black-box approach that uses differential evolution to generate training datasets without prior knowledge of the system. We trained the models using experimental data from 3$\times$3 and 4$\times$4 meshes and deploy the trained networks to control real hardware. We tested on random unitary matrices, our approach outperforms previous methods by $\sim$2 bits of precision. We further extend the method beyond predicting only the squared magnitudes of unitary matrices, as in prior work, to predicting both the amplitude and phase of the applied transformation using a 2$\times$2 gate with coherent detection. This results show that black-box neural network models provide a universal, architecture-independent framework for calibration and control of programmable meshes. 

This work establishes a scalable, architecture-independent route to robust calibration and control of programmable photonic processors, with broad implications for the field. These advances could accelerate the deployment of large-scale photonic systems for quantum computing, optical neural networks, analog photonic computing, microwave photonics, and other applications requiring precise, reconfigurable linear transformations.

\section{Results}
\subsection{Tandem Neural Networks}
\begin{figure}[H]
    \centering
    \includegraphics[width=0.9\linewidth]{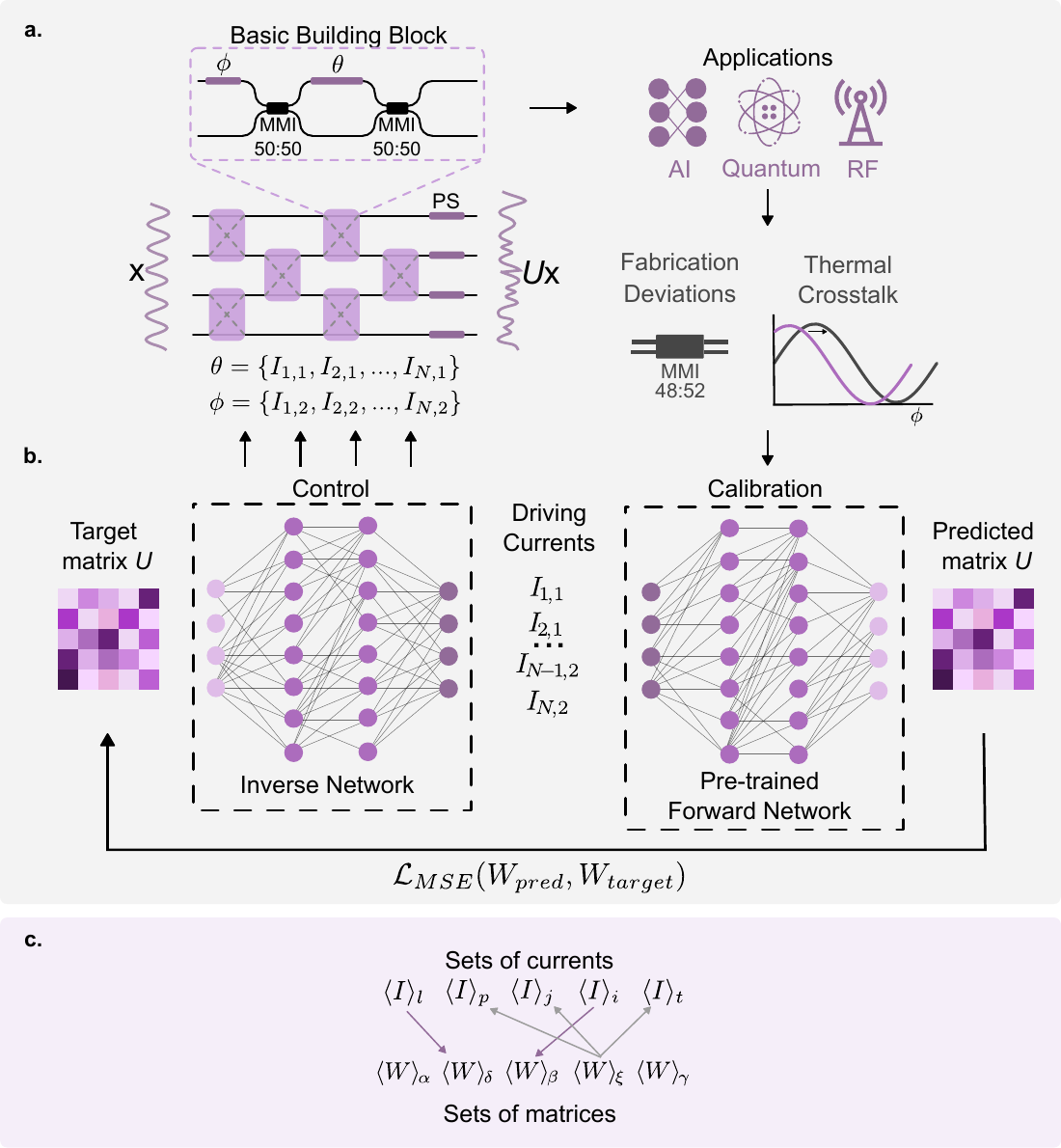}
    \caption{\textbf{a} Rectangular programmable photonic processors, constructed from Mach-Zehnder interferometers (MZIs) with internal and external phase shifters, implement unitary transformations $U$ on input signals $x$. These transformations are the backbone of numerous applications, such as artificial intelligence, quantum computing, and microwave photonics. High performance operation requires specific calibration and control algorithms to capture non-idealities, including fabrication deviations and thermal crosstalk. \textbf{b} Tandem neural networks enable unique calibration and control solutions. First, a forward network learns the hardware calibration (currents-to-matrix mapping). Subsequently, an inverse network is trained against the fixed forward model to predict the required currents for a target matrix. The inverse network is then decoupled to control the processor. \textbf{c} Tandem networks address the many-to-one problem inherent to photonic meshes where distinct phase/current sets yield the same optical response, an ambiguity that limits standard feedforward models.}
    \label{fig:conceptual}
\end{figure}

A direct approach to solving the calibration and control problem, determining the currents required for a given matrix transformation, involves using a single feedforward neural network that maps matrix elements to the necessary currents. However, this method presents scalability challenges due to the data-intensive nature of the task, and the convergence problems during training stemming from the fact that the problem is ill-posed. Specifically, the relationship between current sets and optical matrices in coherent systems is many-to-one, as multiple current configurations can yield the identical optical transformation. This non-uniqueness renders the inverse mapping multivalued, creating ambiguity that hinders standard training and affect its convergence. A schematic representation of this phenomenon is illustrated in Fig. \ref{fig:conceptual}\textbf{c}.

A strictly analogous challenge is encountered in the inverse design of nanophotonic structures \cite{Wiecha:21}. To address this limitation, researchers have proposed the use of Tandem Neural Networks (TNNs) \cite{Yuan2024MultiheadedTN, https://doi.org/10.1002/advs.202401951, He:23, Luo:24, Kim2023DeepArrays, Wu:25}. TNNs employ an architecture similar to autoencoders \cite{Skansi2018}, with the distinction that the latent space represents a physical quantity (currents) and the decoder (forward model) is trained prior to the encoder (inverse model). The architecture comprises an inverse network acting as the encoder and a forward network acting as the decoder, as depicted in Fig. \ref{fig:conceptual}\textbf{b}.  Initially, the forward network learns the direct mapping from applied currents to the resulting linear transformation (calibration). Subsequently, the forward model parameters are frozen, and the inverse model is cascaded with it. The encoder accepts target matrix elements as input and predicts candidate current configurations, which are then fed into the pre-trained decoder to reconstruct the matrix. Training minimizes the loss between the input target matrix and the tandem network's output. Crucially, this strategy resolves the non-uniqueness issue: even if multiple current sets are in principle valid solutions, optimizing in the matrix domain ensures only one of these solutons is selected and hence stable convergence of the training. Upon completion of training, the encoder is decoupled and deployed independently for system control.

\subsection{Photonic Architecture}
Feedforward meshes of tunable MZIs combined with a column of phase shifters (PS) can realize arbitrary unitary linear transformations $U(N)$ using triangular or rectangular layouts, as illustrated in Fig. \ref{fig:conceptual} (green box). The fundamental building block consists of an MZI equipped with an external and an internal phase actuator. For an $N\times N$ transformation, a total of $\frac{N(N-1)}{2}$ MZIs are required. The operation of the (thermo-optic) phase actuators is controlled by applying a current $I_{i,j}$, where $i$ denotes the tunable building block and $j \in \{1,2\}$ corresponds to the external (1) or internal (2) phase shifter. The relationship between the current and the induced phase is given by $\theta = \theta_{0} + \beta I^{2}$, where $\theta_{0}$ represents the initial phase offset arising from fabrication imperfections and $\beta$ denotes the thermo-optic coefficient. Consequently, calibrating the device entails characterizing the $\theta_{0}$ and $\beta$ parameters for all phase actuators within the mesh.

In this work, the rectangular architecture was implemented on a general-purpose photonic integrated processor featuring a hexagonal topology, which is integrated within the iPronics Smartlight system \cite{Perez-Lopez2024}.  Such photonic processors rely on a mesh of Mach–Zehnder interferometers that can be software-configured to define a diverse range of photonic structures, such as filters or tunable delay lines \cite{Perez-Lopez2024}. Furthermore, matrix architectures can be programmed onto this general-purpose hardware, encompassing the optical splitting, vector encoding, linear transformation, and photodetection stages \cite{10.1063/5.0235712}. An optical micrograph of the Smartlight processor is displayed in Fig. \ref{fig:datasets}\textbf{a}, alongside a schematic of the hexagonal mesh in Fig. \ref{fig:datasets}\textbf{b}. Detailed procedures regarding the mapping of the matrix architecture onto the hexagonal topology are provided in Section I of the Supplementary Material.

\subsection{Sampling Methods and Dataset Generation}
\begin{figure}[h]
    \centering
    \includegraphics[width=1.0\linewidth]{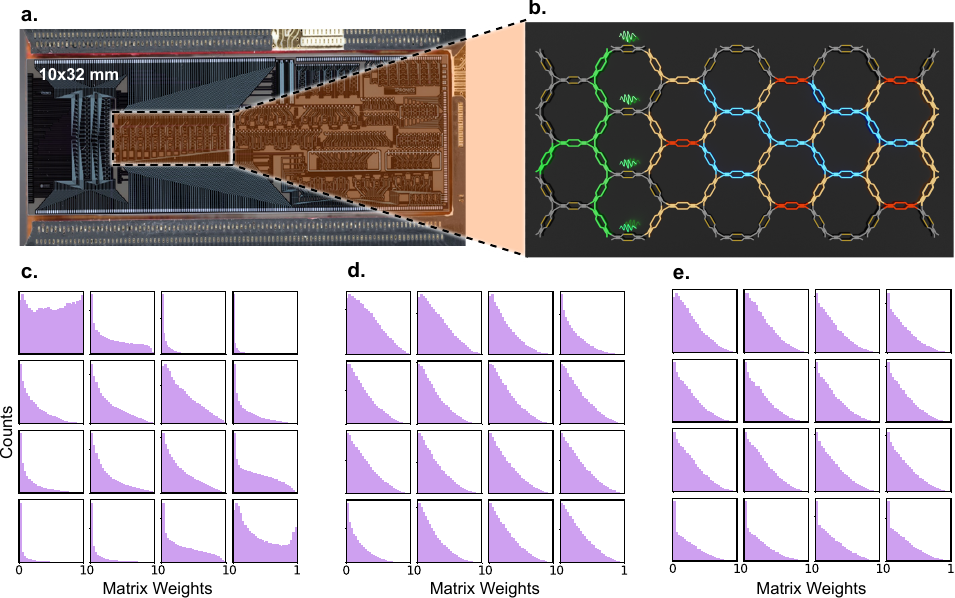}
    \caption{\textbf{a} Optical micrograph of the Smartlight processor, \textbf{b} schematic of the hexagonal mesh included in the hexagonal processor and the configuration required for the implementation of a rectangular topology for matrix multiplication. Green MZIs are used to build a $1\times4$ splitter tree, the orange MZIs represent the \textit{cross} state while the red MZIs represent the \textit{bar}, and the blue MZIs represent the tunable elements used for matrix multiplications. \textbf{c-d-e} represent the dataset used for the training of the $4\times4$ models using the uniform, architecture aware, and optimized samplings, respectively.}
    \label{fig:datasets}
\end{figure}

Training neural network models necessitates a comprehensive dataset mapping the sets of currents applied to the phase actuators to their corresponding optical matrices. Experimentally, phase shifters were configured according to sampled currents, and the resulting matrices were characterized by injecting identity states via the input MZI array (Fig. \ref{fig:datasets}\textbf{b}) and recording the response on the integrated photodetectors. Developing a robustly generalizable model requires a training dataset that accurately reflects the probability distribution of the target application, in this case, the full unitary group. 

Prior studies have typically generated datasets by sampling currents from a uniform distribution that covers the corresponding phase range $[0, \pi]$ \cite{Cem:23, 10416579, Fan:25, 10.1063/5.0293082}. However, uniform sampling of currents (or phases) in coherent architectures, whether rectangular or triangular, fails to yield a uniform distribution over the unitary space \cite{PhysRevApplied.11.064044, Russell2017}. Instead, this method inherently biases the generated data toward banded unitary matrices. Consequently, during training, models are exposed to a restricted subspace of the unitary group. This severely compromises the model's generalization capability: accuracy degrades significantly when inferencing arbitrary random unitary matrices residing in part of the under sampled region of the unitary space. This limitation has been largely overlooked in prior works, as evaluations frequently relied on test sets drawn from the same biased distribution as the training data. The skewed distribution of matrix elements resulting from uniformly sampled currents for a $4\times4$ mesh is illustrated in Fig. \ref{fig:datasets}\textbf{c}. In the subsequent section, we quantify this performance degradation when testing on unbiased random unitary matrices and introduce two novel approaches to resolve this sampling bottleneck.

\subsubsection{Architecture-Aware Sampling (AAS)}
To achieve a uniform distribution of matrices within the unitary space (Haar measure), the currents applied to the MZIs must adhere to a specific distribution derived from random matrix theory \cite{PhysRevApplied.11.064044}, whereas the external phase shifters may be sampled uniformly. This MZI current distribution is governed by a sensitivity index $\alpha_{nl}$, which depends on the spatial position $(n,l)$ of the phase shifter within the mesh. For a rectangular topology, the required current-phase relationship is expressed as:
\begin{equation}
    I^{2}_{nl} = \frac{2\arccos\left(\sqrt[\uproot{2}  2\alpha_{nl}]{\xi_{nl}}\right) - \theta_{0, nl}}{\beta},
    \label{eq:current}
\end{equation}
where $\beta$ represents the thermo-optic coefficient, $\theta_{0, nl}$ denotes the initial phase offset arising from fabrication imperfections, and $\xi_{nl}$ is a random variable uniformly distributed in the interval $[0, 1]$. A rigorous derivation of this distribution is provided in the Methods section.

The coefficient $\beta$ can be approximated as $\frac{\pi R_{0}}{P_{\pi}}$, with $R_{0}$ being the heater resistance and $P_{\pi}$ the power required to induce a $\pi$ phase shift. Consequently, the set of initial phase offsets $\theta_{0, nl}$ constitutes the sole unknown variable. We estimated these values using a differential evolution algorithm. In each generation, a batch of $500$ samples was generated, and the algorithm optimized the phase offsets to minimize the Wasserstein distance between the generated matrix distribution and the ideal distribution of the square modulus of unitary matrices. The resulting distribution for the $4\times4$ case is visualized in Fig. \ref{fig:datasets}\textbf{d}. Further details regarding the evolutionary algorithm and the corresponding $3\times3$ dataset are available in Section II of the Supplementary Material.

\subsubsection{Optimized Sampling (OS)}
While Architecture-Aware Sampling (AAS) proves effective, it necessitates a priori knowledge of the photonic processor’s topology, effectively functioning as a pre-calibration stage. To circumvent this dependency, we propose an Optimized Sampling (OS) strategy. This approach initiates by defining an ideal target distribution comprising the squared moduli of $N$ randomly generated unitary matrices. Subsequently, for each element in this synthetic dataset, a differential evolution algorithm optimizes the set of driving currents required to experimentally realize the specific target matrix on the hardware by minimizing the correlation between the experimental and target matrices. This process is iterated until the complete target dataset is successfully mapped to the chip. The resulting optimized distribution for the $4 \times 4$ case is visualized in Fig. \ref{fig:datasets}\textbf{e}. Comprehensive details regarding the optimization algorithm and a comparative analysis of the ideal versus optimized datasets for both $3 \times 3$ and $4 \times 4$ architectures are provided in Section II of the Supplementary Material.

\subsection{Model training and evaluation} 
\subsubsection*{3x3 Processor}
\begin{figure}[H]
    \centering
    \includegraphics[width=0.7\linewidth]{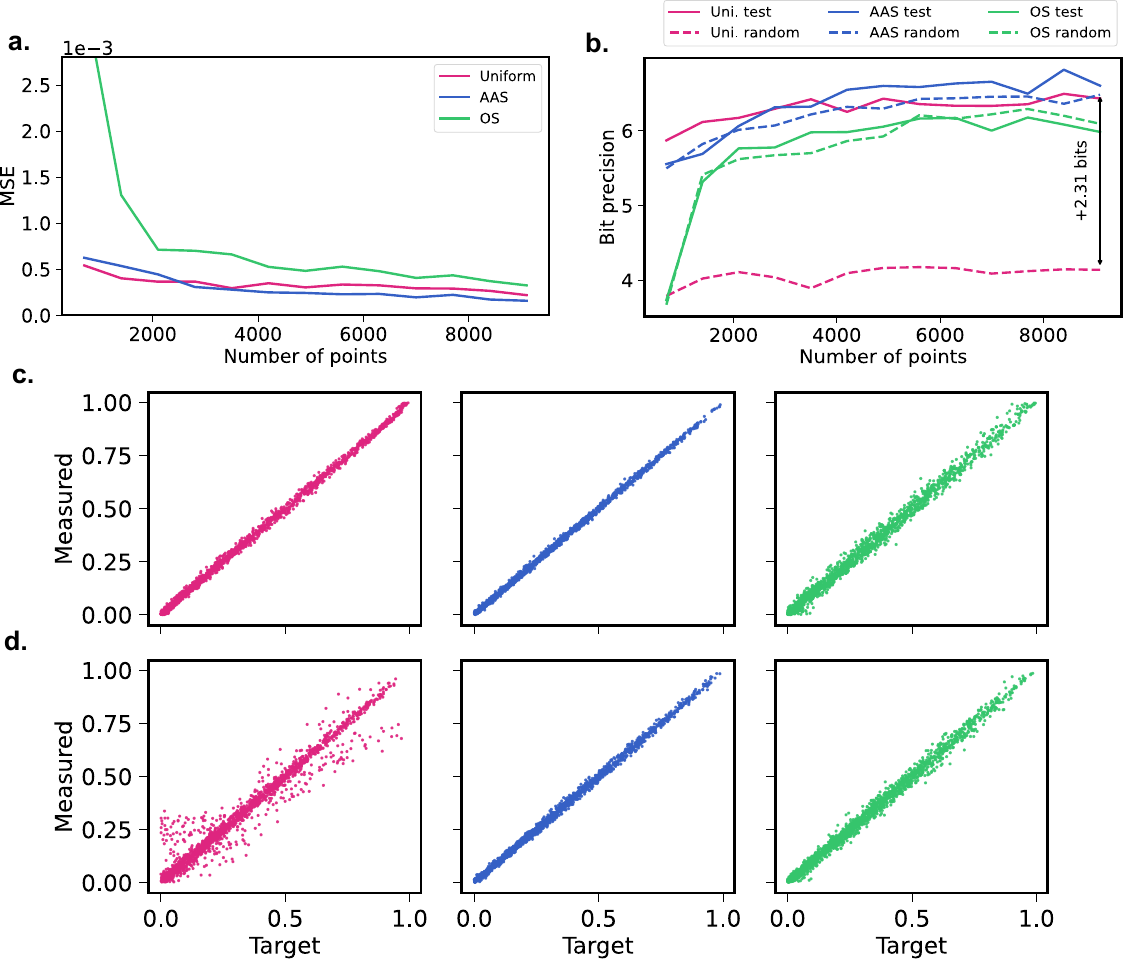}
    \caption{\textbf{a} MSE evolution of the optimal models as a function of training dataset size. \textbf{b} Experimental bit precision measured on the distribution-matched set (solid lines) and random unitary matrices (dotted lines). \textbf{c} Comparison between target and on-chip measured matrix weights evaluated on the distribution-matched set. \textbf{d} Comparison between target and measured weights evaluated on random unitary matrices for the $3\times3$ processor.}
    \label{fig:results3_3}
\end{figure}

For the $3\times3$ processor, the available training dataset comprised a maximum of $13000$ samples. To assess data efficiency, we conducted a scaling study by training the model on subset sizes ranging from $1000$ to $13000$ in increments of $1000$. Each subset was divided into 70$\%$ for training, 15$\%$ for validation, and 15$\%$ for testing. For statistical robustness, each configuration was trained over $10$ independent runs. The evolution of the best MSE on the test set for the uniform, architecture aware and optimized sampling techniques is depicted in Fig.~\ref{fig:results3_3}\textbf{a}. We observed that all three strategies tended to converge beyond $6000$ training samples, although the OS required a larger dataset to reach convergence compared to the Uniform and AAS approaches.

Following training, the optimal models for each dataset size were experimentally validated on the physical hardware. This evaluation utilized two distinct datasets: a distribution-matched set of $500$ points (consistent with the training distribution) and a set of $500$ randomly generated unitary matrices. Figure~\ref{fig:results3_3}\textbf{b} illustrates the results, distinguishing between the distribution-matched set (solid lines) and the random matrices (dotted lines) in terms of experimental bit precision calculated as $P_{\text{bit}} = \log_{2} \left( \frac{\max(w) - \min(w)}{\sigma_{\epsilon}} \right)$. Notably, while the AAS and OS techniques demonstrated robust performance across both scenarios, the uniform sampling approach suffered significant degradation when evaluated on random matrices. Consequently, in the best-case scenario, our AAS solution yielded a precision improvement of $2.31$ bits on random matrices compared to previously proposed neural network solutions based on uniform sampling, while the OS solution provided a $1.9$ bit improvement. A comparison between the target and on-chip measured matrix weights for both the distribution-matched and random unitary matrices is depicted in Fig.~\ref{fig:results3_3}\textbf{c-d}.

\subsubsection*{4x4 Processor}
\begin{figure}[H]
    \centering
    \includegraphics[width=0.7\linewidth]{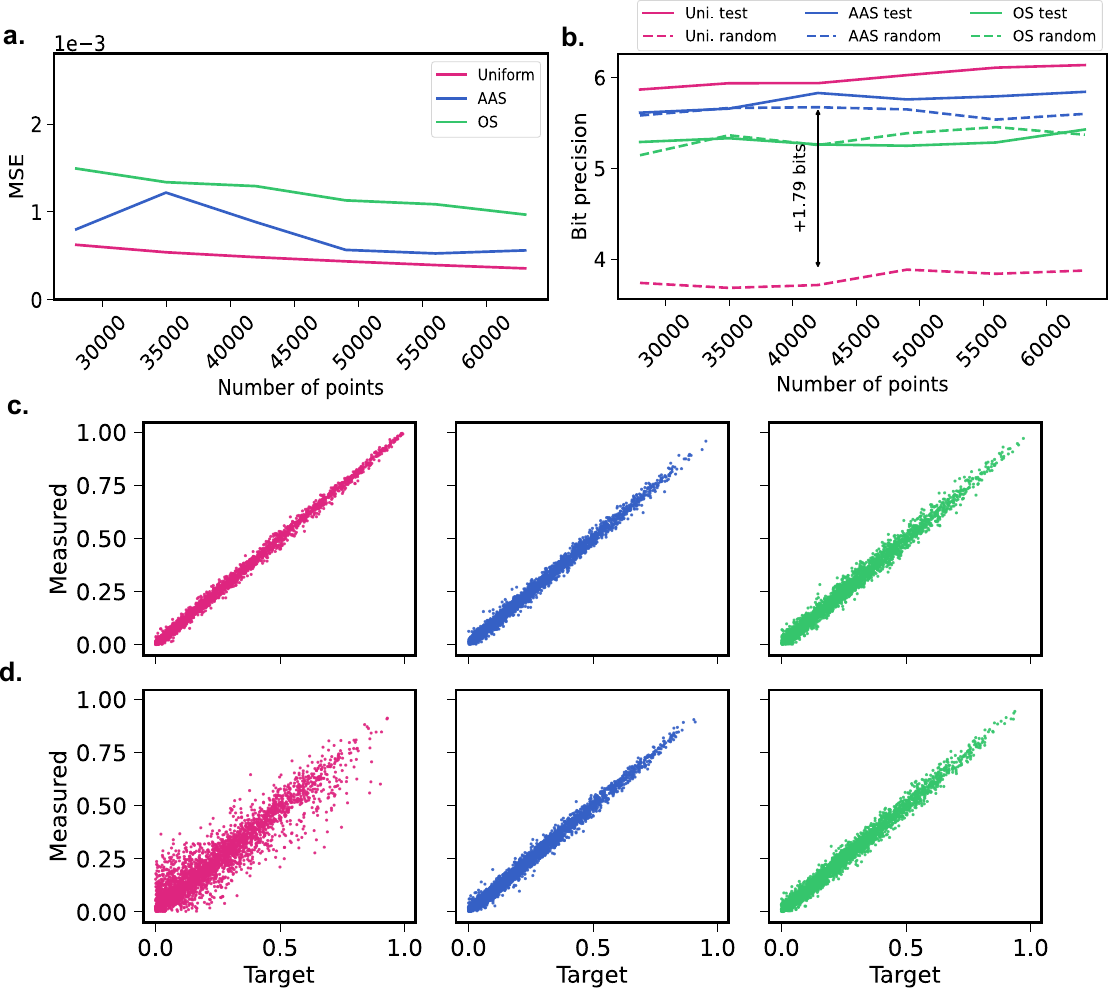}
    \caption{\textbf{a} MSE evolution of the optimal models as a function of training dataset size. \textbf{b} Experimental bit precision measured on the distribution-matched set (solid lines) and random unitary matrices (dotted lines). \textbf{c} Comparison between target and on-chip measured matrix weights evaluated on the distribution-matched set. \textbf{d} Comparison between target and measured weights evaluated on random unitary matrices for the $4\times4$ processor.}
    \label{fig:results4_4}
\end{figure}

For the $4\times4$ processor, the dataset consisted of $90000$ samples. To mitigate the latency constraints imposed by the control electronics and accelerate the data generation process, we developed a digital twin of the photonic processor. This simulation environment, incorporating experimentally derived calibration parameters, was utilized to execute the Architecture-Aware and Optimized Sampling protocols and obtained the required currents to generate the datasets. Then, the currents were applied on the physical device to obtain the final experimental datasets. The validation and testing of the neural network models were also performed directly on the physical hardware. A comprehensive analysis regarding the data acquisition time is presented in the Discussion Section. Models were trained on data subsets starting from $40000$ samples in increments of $10000$. Each subset was divided into 70$\%$ for training, 15$\%$ for validation, and 15$\%$ for testing. To ensure statistical significance, each configuration underwent $10$ independent training runs. The evolution of the MSE on the test set is depicted in Fig.~\ref{fig:results4_4}\textbf{a}. In this scenario, the AAS method consistently outperformed OS, whereas uniform sampling yielded the lowest numerical MSE. However, the experimental validation in terms of bit precision, presented in Fig.~\ref{fig:results4_4}\textbf{b}, mirrors the trends observed in the $3\times3$ case. These results corroborate the significant performance degradation of uniform sampling when applied to random matrices, in contrast to the robustness of the AAS and OS techniques. Specifically, the optimal AAS and OS models achieved precision improvements of $1.79$ and $1.58$ bits, respectively, over the uniform baseline on random matrices. Notably, the minimal deviation between the precisions obtained on the distribution-matched set and the randomly generated matrices highlights the universality and superior generalization capability of our proposed solutions. A comparison between the target and on-chip measured matrix weights for both the distribution-matched and random unitary matrices is depicted in Fig.~\ref{fig:results4_4}\textbf{c-d}.

\subsection{Analog Programmable-Photonic and Quantum Computing}
\begin{figure}[H]
    \centering
    \includegraphics[width=0.95\linewidth]{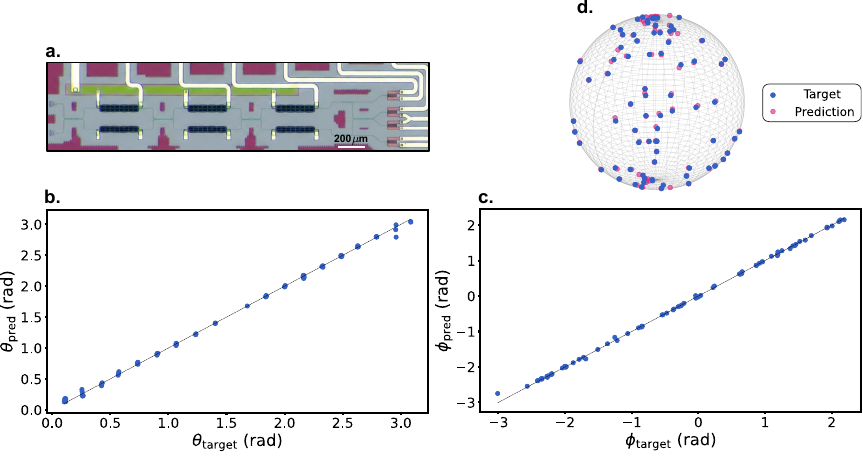}
    \caption{\textbf{a} Optical micrograph of the on-chip universal gate employed for arbitrary state rotation. \textbf{b} Comparison between target and predicted $\theta$ rotations. \textbf{c} Comparison between target and predicted $\phi$ rotations. \textbf{d} Visualization of the target and predicted optical states on the Bloch sphere.}
    \label{fig:phases}
\end{figure}

Analog programmable-photonic computing \cite{https://doi.org/10.1002/lpor.202200360, Macho-Ortiz2026AnalogInformation, lopezmarch2026noiseanalogprogrammablephotoniccomputation} and photonic quantum computing \cite{Zhou2025AComputing} rely on the implementation of gates for the mathematical rotation of information units. A critical component of these architectures is the detection stage, which necessitates the processing of not only optical power but rather signal amplitude and phase. To demonstrate the general applicability of our solution for the calibration and control of programmable photonic circuits, we evaluated its use on a $2\times2$ universal gate employing a coherent detection system, as depicted in Fig.~\ref{fig:phases}\textbf{a}. The dataset comprised $400$ samples, partitioned into $70\%$ for training, $15\%$ for validation, and $15\%$ for testing. Each sample contained the driving currents for the internal MZI phase shifter and the two external phase shifters at the input and output. The objective was to predict the rotation angles $\theta$ and $\phi$ imparted to the optical state on a unit-radius Bloch sphere. Network optimization followed the protocol described in the previous experiments. The test set predictions for $\theta$ and $\phi$, presented in Fig.~\ref{fig:phases}\textbf{b-c}, demonstrate a excellent agreement between the target and predicted rotations. These results are visualized on the Bloch sphere in Fig.~\ref{fig:phases}\textbf{d}. To the best of our knowledge, this constitutes the first experimental demonstration of data-driven black-box models for phase prediction in programmable photonic circuits, paving the way for scalable calibration and control algorithms in coherent photonic and quantum computing.  

\subsection{Photonic Neural Networks}
\begin{figure}[H]
    \centering
    \includegraphics[width=0.6\linewidth]{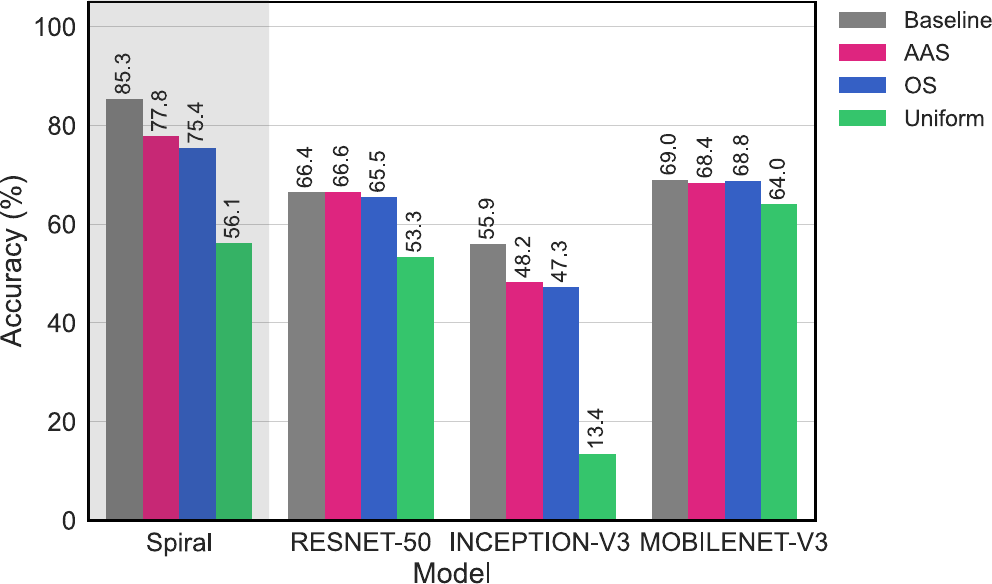}
    \caption{Test set accuracy comparison for various neural network architectures on the CIFAR-10 dataset. Performance is evaluated using bit precision derived from AAS, OS, and uniform sampling techniques, compared against a 32-bit precision baseline. The `Spiral' entry denotes a feedforward network trained on a spiral dataset, where the weight matrices were experimentally tested directly on the photonic processor.}
    \label{fig:pnn_comparative}
\end{figure}

Another promising application of photonic processors is their use in neural networks, particularly as matrix accelerators during inference. To demonstrate the advantage of our proposed method over previous works based on uniform sampling, we trained a four-layer feedforward neural network using a spiral dataset, a non-linear classification dataset that consists of multiple spiral-shaped arms, each representing a different class. The weight matrices of the best models for each sampling technique were implemented on the photonic processor to compare their performance on the test set. The results are presented in Fig.~\ref{fig:pnn_comparative}, where it can be observed that the performance degradation using the AAS and OS methods is limited to 8\% when compared to a 32-bit precision digital model, whereas uniform sampling results in a degradation of nearly 20\%. Subsequently, we trained three architectures, ResNet-50, Inception-V3, and MobileNet-V3, using the CIFAR-10 dataset and compared their test set accuracies when the matrices were reduced to the precision levels obtained with the studied sampling techniques. These results are shown in Fig.~\ref{fig:pnn_comparative}, where uniform sampling consistently exhibits the lowest performance compared to the full-precision baseline. In contrast, AAS and OS demonstrate similar performance with minimal degradation.

\section{Discussion}
In this work, we demonstrate the use of neural networks for the calibration and control of photonic integrated processors. In coherent architectures, distinct sets of control signals can yield identical optical responses, creating a many-to-one problem. This characteristic imposes additional computational overhead on the training of neural networks. To address this challenge, we propose employing tandem neural networks, a solution previously validated in the field of inverse design for photonic structures. The use of tandem neural networks divides the training into two phases: first, a network learns the mapping between the applied currents and the optical response (calibration phase), subsequently, a second network is cascaded with the calibration network to learn the currents required for a given optical response (control phase). This tandem configuration significantly enhances the learning performance of the system.

A second critical aspect addressed in this study is the generation of the training dataset. Previous works rely on the use of a uniform sampling technique, which, in the case of photonic processors, yields a skewed optical response that does not represent the distribution of matrix elements in the unitary space. To overcome this limitation, we propose two distinct techniques. The first, Architecture-Aware Sampling (AAS), is derived from random matrix theory. We utilize the current-phase relationship inherent to coherent architectures and determine the required parameters using a differential evolution algorithm that minimizes the Wasserstein distance between the on-chip generated matrices and the ideal distribution in the unitary space. The second sampling technique, Optimized Sampling (OS), requires no prior knowledge of the system and is based on defining an ideal set of target unitary matrices. Subsequently, a differential evolution algorithm is employed to find the required currents to generate each of these matrices experimentally.

We validated these sampling techniques on $3\times3$ and $4\times4$ matrix processors employing a rectangular topology and benchmarked them against the standard uniform sampling technique. Each model was trained using varying dataset sizes. Post-training, we evaluated the models on two distinct test sets. The first test set followed the same distribution as the training data. In this scenario, uniform sampling yielded results comparable to, and in the $4\times4$ case superior to, the proposed approaches, achieving bit precisions of 6.3 (uniform), 6.9 (AAS), and 6.0 (OS) for the $3\times3$ processor, and 6.15, 5.8, and 5.35, respectively, for the $4\times4$ processor. However, regarding the second test set, which comprised matrices randomly sampled from the full unitary space, the performance of the uniform sampling models dropped to approximately 4 bits for both processors. In contrast, the AAS and OS-based models maintained consistent performance. This significant drop in bit precision demonstrates that relying solely on uniform sampling fails to accurately represent the full scope of the control problem required for universal generalization.

A primary limitation of the current experimental setup lies in the bandwidth constraints of the electronic readout system. The current instrumentation limits the measurement throughput to approximately $0.54\,\text{s}$ and $0.72\,\text{s}$ per matrix for the $3\times3$ and $4\times4$ processors, respectively. Consequently, the data acquisition for the Architecture-Aware approach required $45\,\text{h}$ ($3\times3$) and $300\,\text{h}$ ($4\times4$), while the Optimized approach extended to $162\,\text{h}$ ($3\times3$) and $1,600\,\text{h}$ ($4\times4$). 

However, this bottleneck can be circumvented by integrating high-speed control electronics. By operating within the thermal time constant of the phase shifters ($300\,\mu\text{s}$ rise/fall time), the acquisition time could be reduced by a factor of $1000$. Furthermore, the integration of high-speed electro-optic actuators (operating in the GHz regime) would drastically further reduce these durations. Finally, it is important to note that this extensive data acquisition represents a one-time calibration expense. Moreover, the required dataset size can be substantially decreased in future iterations by employing transfer learning techniques \cite{Zibar2023AddressingLearning}.

Beyond intensity detection, we demonstrate that our proposal can predict the optical phase using a $2\times2$ universal gate with coherent detection. This setup yielded near-perfect prediction of rotations for a 2-D unit of information in terms of $\theta$ and $\phi$ angles, paving the way for the control of photonic circuits in quantum and analog computing. To validate the practical utility of our solution, we applied it to photonic neural networks by training a simple feedforward network on the Spiral dataset. We compared the accuracy on the test set when the weight matrices were generated directly on-chip using our trained control models. The results indicate that while the AAS and OS methods exhibit minimal accuracy degradation, the uniform sampling approach suffers from a performance drop exceeding $20\%$. Finally, we trained three distinct architectures on the CIFAR-10 dataset and evaluated their performance by injecting Gaussian noise corresponding to the measured bit precision of the trained models. Once again, our proposal outperforms uniform sampling, particularly in the Inception-V3 architecture, where our solution shows a degradation of only $\approx 7\%$ compared to the full-precision baseline, whereas uniform sampling results in a degradation exceeding $40\%$.

While previous studies have addressed the calibration and control of photonic processors using purely neural network-based approaches, to the best of our knowledge, our work is the first to provide a universal solution capable of implementing any unitary matrix on the system. Furthermore, we have demonstrated architecture independence and complex-valued weight prediction capabilities. In terms of computational efficiency, our solution achieves higher on-chip accuracies than prior methods while maintaining a comparable number of network parameters and dataset sizes. A comprehensive comparison between our proposal and the existing literature is presented in Table~\ref{tab:comparison}.

\begin{sidewaystable}[htpb]
    \centering
    \begin{threeparttable}
        \caption{\title{Experimental calibration and control of photonic integrated circuits using exclusively neural networks.}} 
        \label{tab:comparison}
        \begin{tabular}{ccccccccc}
            \toprule
            \textbf{Ref.} & 
            \makecell{\textbf{Problem} \\ \textbf{Solved}} & 
            \makecell{\textbf{Processor} \\ \textbf{Size}} &
            \makecell{\textbf{Controlled} \\ \textbf{Signals}} & 
            \makecell{\textbf{Dataset} \\ \textbf{Size}} &
            \makecell{\textbf{Network} \\ \textbf{Parameters}} & 
            \makecell{\textbf{On-chip} \\ \textbf{Bit-Precision}} & 
            \makecell{\textbf{Universal} \\ \textbf{Solution}} & 
            \makecell{\textbf{Complex Weight} \\ \textbf{Prediction}}\\
            \midrule
             \cite{PhysRevApplied.15.044003} & Calibration & 3$\times$3 & 2 & 2400 & 83804 & - & No & No\\
            \midrule
             \cite{Cem:23} & Calibration & 3$\times$3 & 9 & 3700 & 13769 & 5.8\tnote{[1]} & No & No \\
            \midrule
             \cite{Youssry2024ExperimentalControl} & Calibration & 3$\times$3 & 4 & 7000 & 7300 & - & No & No\\
            \midrule
            \cite{Fan:25} & \makecell{Calibration \\ and Control} & 4$\times$4 & 16\tnote{[2]} & 64000 & $3.3 \times 10^{6}$ & -\tnote{[3]} & No & No \\
            \midrule
            \cite{10.1063/5.0293082} & \makecell{Calibration \\ and Control} & 3$\times$3 & 3 & $10^{6}$ & $5.2 \times 10^{5}$ & 2.8\tnote{[1]} & No & No \\
            \midrule
            \makecell{\textbf{This} \\ \textbf{work}} & \makecell{Calibration \\ and Control} & 3$\times$3 & 6 & 8000 & $1.2 \times 10^{6}$ & \textbf{6.4} & \textbf{Yes} & \textbf{Yes} \\
            \midrule
            \makecell{\textbf{This} \\ \textbf{work}} & \makecell{Calibration \\ and Control} & 4$\times$4 & 12 & 45000 & $5.0 \times 10^{6}$ & \textbf{5.6} & \textbf{Yes} & \textbf{Yes}\tnote{[4]} \\
            \bottomrule
        \end{tabular}
        
        \begin{tablenotes}
            \small 
            \item[1] Calculated on the test set of the dataset, not directly on-chip.
            \item[2] Four signals correspond to the output phase shifter array, which has no influence on the result due to the use of single photodetectors.
            \item[3] Experimental fidelity of 0.989 using cosine similarity.
            \item[4] Demonstrated on a 2$\times$2 universal gate.
        \end{tablenotes}
    \end{threeparttable} 
\end{sidewaystable}

\section{Methods}
\label{sec:Methods}
\subsection{Experimental set up}
Data generation for the $3\times3$ and $4\times4$ experiments was performed using the Smartlight processor, employing its native control electronics and integrated photodetectors. The optical source consisted of a continuous-wave (CW) laser (Aerodiode) operating at $1550$~nm with an output power of $10$~dBm. The $2\times2$ gate experiment utilized a silicon-on-insulator (SOI) chip fabricated by the AMF SiP foundry, featuring thermo-optic phase shifters and integrated Germanium (Ge) photodetectors. This device was driven by multi-channel current sources (Qontrol Ltd.), while the photodetector outputs were acquired using four Keithley 2400 Source Measure Units (SMUs). All computational training was executed on an NVIDIA GeForce RTX 2070 SUPER GPU. 

\subsection{Derivation of the AAS current distribution}
\label{sec:aas}
Uniform sampling of the phases in a coherent programmable photonic processor does not yield a uniform distribution over the space of unitary matrices (Haar measure) \cite{PhysRevApplied.11.064044}. To determine the required sampling distribution, we derive the expression from the transmission coefficient of the Mach-Zehnder Interferometer (MZI), given by:
\begin{equation}
    t_{nl} = \sqrt[\uproot{2} \alpha_{nl}]{\xi_{nl}},
\end{equation}
where $\alpha_{nl}$ is a position-dependent sensitivity index within the mesh, and $\xi_{nl}$ is a random variable uniformly distributed in the interval $[0, 1]$. Based on the MZI transfer matrix, the required phase shift is defined as:
\begin{equation}
    \theta_{nl} = 2\arccos\left(\sqrt[\uproot{2}  2\alpha_{nl}]{\xi_{nl}}\right).
\end{equation}
By incorporating the constitutive relationship of thermo-optic phase shifters, where the phase shift relates to the applied current $I$ and a bias $\theta_{0,nl}$, we obtain the required current distribution:
\begin{equation}
\begin{split}
    \theta_{nl} &= \theta_{0,nl} + \beta I^{2}_{nl} = 2\arccos\left(\sqrt[\uproot{2}  2\alpha_{nl}]{\xi_{nl}}\right) \\
    I^{2}_{nl} &= \frac{2\arccos\left(\sqrt[\uproot{2}  2\alpha_{nl}]{\xi_{nl}}\right) - \theta_{0, nl}}{\beta}.
\end{split}
\end{equation}

\subsection{Training of the models}
The training strategy employed a tandem network architecture. Initially, the forward model was trained to map current distributions to the target matrices by minimizing the mean squared error (MSE). To accurately reflect the physics of thermo-optic phase shifters, squared currents were utilized as inputs rather than raw values. Subsequently, the inverse model was cascaded with the pre-trained forward model. In this stage, the inverse model's weights were optimized by minimizing the reconstruction MSE of the full tandem network, while the forward model parameters remained frozen. All currents and weights were normalized to the interval $[-1, 1]$, and a hyperbolic tangent activation function was deployed at the final layer of each model. Hyperparameter tuning was conducted using the Optuna framework \cite{akiba2019optunanextgenerationhyperparameteroptimization}. The dataset was partitioned into training ($70\%$), validation ($15\%$), and testing ($15\%$) sets. Finally, optimization was performed using the Adam algorithm, incorporating an early stopping protocol that terminated training if the validation MSE failed to improve by a threshold of $0.001$ over $50$ consecutive epochs.

\subsection*{Acknowledgments}
This work was supported by the European Research Council (ERC) Advanced Grant program under grant agreement No. 101097092 (ANBIT), the ERC Starting Grant program under grant agreement No. 101076175 (LS-Photonics Project) and grant agreement No. 101041131 (BEAMS project), and by the ‘National Hub of Excellence in Quantum Communications’ initiative of the Ministry for Digital Transformation and Public Administration with NextGenerationEU funds within the framework of the Recovery, Transformation and Resilience Plan and the Recovery Mechanism.

\subsection*{Author Contributions}
J.R.R.C., D.P.L. and D.M. conceived the idea of data-driven control of programmable photonic circuits, J.R.R.C. and D.M. developed the algorithms. J.R.R.C. conducted the simulations and designed the experiments with input from D.M. and B.S. J.R.R.C performed the experiments and analyzed the data. J.R.R.C. led the writing of the manuscript, with contributions from D.M., B.S., D.P.L., and J.C.F., and J.C.F. supervised the study.

\subsection*{Conflicts of Interest}
The authors declare no conflict of interest.

\subsection*{Data Availability}
Data are available from the corresponding author upon reasonable request.

\printbibliography{}

@article{Prabhu2020,
    title = {{Accelerating recurrent Ising machines in photonic integrated circuits}},
    year = {2020},
    journal = {Optica},
    author = {Prabhu, Mihika and Roques-Carmes, Charles and Shen, Yichen and Harris, Nicholas and Jing, Li and Carolan, Jacques and Hamerly, Ryan and Baehr-Jones, Tom and Hochberg, Michael and {\v{C}}eperi{\'{c}}, Vladimir and Joannopoulos, John D and Englund, Dirk R and Solja{\v{c}}i{\'{c}}, Marin},
    number = {5},
    pages = {551--558},
    volume = {7},
    publisher = {Optica Publishing Group},
    url = {http://opg.optica.org/optica/abstract.cfm?URI=optica-7-5-551},
    doi = {10.1364/OPTICA.386613}
}

@article{Dhand_2016,
    title = {{Accurate and precise characterization of linear optical interferometers}},
    year = {2016},
    journal = {Journal of Optics},
    author = {Dhand, Ish and Khalid, Abdullah and Lu, He and Sanders, Barry C},
    number = {3},
    month = {2},
    pages = {35204},
    volume = {18},
    publisher = {IOP Publishing},
    url = {https://dx.doi.org/10.1088/2040-8978/18/3/035204},
    doi = {10.1088/2040-8978/18/3/035204}
}

@article{https://doi.org/10.1002/lpor.202200360,
    title = {{Analog Programmable-Photonic Computation}},
    year = {2023},
    journal = {Laser {\&} Photonics Reviews},
    author = {Macho-Ortiz, Andrés and P{\'{e}}rez-L{\'{o}}pez, Daniel and Aza{\~{n}}a, José and Capmany, José},
    number = {10},
    pages = {2200360},
    volume = {17},
    url = {https://onlinelibrary.wiley.com/doi/abs/10.1002/lpor.202200360},
    doi = {https://doi.org/10.1002/lpor.202200360}
}

@incollection{Skansi2018,
    title = {{Autoencoders}},
    year = {2018},
    booktitle = {Introduction to Deep Learning: From Logical Calculus to Artificial Intelligence},
    author = {Skansi, Sandro},
    pages = {153--163},
    publisher = {Springer International Publishing},
    url = {https://doi.org/10.1007/978-3-319-73004-2_8},
    address = {Cham},
    isbn = {978-3-319-73004-2},
    doi = {10.1007/978-3-319-73004-2{\_}8}
}

@inproceedings{9975587,
    title = {{Automatic Self-calibration of Programmable Photonic Processors}},
    year = {2022},
    booktitle = {2022 IEEE Photonics Conference (IPC)},
    author = {L{\'{o}}pez-Hern{\'{a}}ndez, Aitor and Guti{\'{e}}rrez-Zubillaga, Mikel and P{\'{e}}rez-L{\'{o}}pez, Daniel},
    pages = {1--2},
    doi = {10.1109/IPC53466.2022.9975587}
}

@article{Alexiev2021,
    title = {{Calibrating rectangular interferometer meshes with external photodetectors}},
    year = {2021},
    journal = {OSA Continuum},
    author = {Alexiev, Christopher and Mak, Jason C C and Sacher, Wesley D and Poon, Joyce K S},
    number = {11},
    pages = {2892--2904},
    volume = {4},
    publisher = {Optica Publishing Group},
    url = {https://opg.optica.org/osac/abstract.cfm?URI=osac-4-11-2892},
    doi = {10.1364/OSAC.437918}
}

@article{PhysRevApplied.15.044003,
    title = {{Calibration of Multiparameter Sensors via Machine Learning at the Single-Photon Level}},
    year = {2021},
    journal = {Phys. Rev. Appl.},
    author = {Cimini, Valeria and Polino, Emanuele and Valeri, Mauro and Gianani, Ilaria and Spagnolo, Nicolò and Corrielli, Giacomo and Crespi, Andrea and Osellame, Roberto and Barbieri, Marco and Sciarrino, Fabio},
    number = {4},
    pages = {44003},
    volume = {15},
    publisher = {American Physical Society},
    url = {https://link.aps.org/doi/10.1103/PhysRevApplied.15.044003},
    doi = {10.1103/PhysRevApplied.15.044003}
}

@article{8610179,
    title = {{Canceling Thermal Cross-Talk Effects in Photonic Integrated Circuits}},
    year = {2019},
    journal = {Journal of Lightwave Technology},
    author = {Milanizadeh, Maziyar and Aguiar, Douglas and Melloni, Andrea and Morichetti, Francesco},
    number = {4},
    pages = {1325--1332},
    volume = {37},
    doi = {10.1109/JLT.2019.2892512}
}

@article{He:23,
    title = {{Constrained tandem neural network assisted inverse design of metasurfaces for microwave absorption}},
    year = {2023},
    journal = {Opt. Express},
    author = {He, Xiangxu and Cui, Xiaohan and Chan, C T},
    number = {24},
    month = {11},
    pages = {40969--40979},
    volume = {31},
    publisher = {Optica Publishing Group},
    url = {https://opg.optica.org/oe/abstract.cfm?URI=oe-31-24-40969},
    doi = {10.1364/OE.506936}
}

@article{Cem:23,
    title = {{Data-Driven Modeling of Mach-Zehnder Interferometer-Based Optical Matrix Multipliers}},
    year = {2023},
    journal = {J. Lightwave Technol.},
    author = {Cem, Ali and Yan, Siqi and Ding, Yunhong and Zibar, Darko and Ros, Francesco Da},
    number = {16},
    pages = {5425--5436},
    volume = {41},
    publisher = {Optica Publishing Group},
    url = {https://opg.optica.org/jlt/abstract.cfm?URI=jlt-41-16-5425}
}

@article{10.1063/5.0293082,
    title = {{Data-driven programming of feedforward photonic integrated circuits}},
    year = {2026},
    journal = {APL Photonics},
    author = {Cavicchioli, G and Masini, G and Sances, F M and Morichetti, F and Melloni, A},
    number = {1},
    pages = {10801},
    volume = {11},
    url = {https://doi.org/10.1063/5.0293082},
    doi = {10.1063/5.0293082},
    issn = {2378-0967}
}

@article{Wiecha:21,
    title = {{Deep learning in nano-photonics: inverse design and beyond}},
    year = {2021},
    journal = {Photon. Res.},
    author = {Wiecha, Peter R and Arbouet, Arnaud and Girard, Christian and Muskens, Otto L},
    number = {5},
    month = {5},
    pages = {B182--B200},
    volume = {9},
    publisher = {Optica Publishing Group},
    url = {https://opg.optica.org/prj/abstract.cfm?URI=prj-9-5-B182},
    doi = {10.1364/PRJ.415960}
}

@article{Shen2017,
    title = {{Deep learning with coherent nanophotonic circuits}},
    year = {2017},
    journal = {Nature Photonics},
    author = {Shen, Yichen and Harris, Nicholas C and Skirlo, Scott and Prabhu, Mihika and Baehr-Jones, Tom and Hochberg, Michael and Sun, Xin and Zhao, Shijie and Larochelle, Hugo and Englund, Dirk and Solja{\v{c}}i{\'{c}}, Marin},
    number = {7},
    pages = {441--446},
    volume = {11},
    url = {https://doi.org/10.1038/nphoton.2017.93},
    doi = {10.1038/nphoton.2017.93},
    issn = {1749-4893}
}

@article{SeyedinNavadeh2024,
    title = {{Determining the optimal communication channels of arbitrary optical systems using integrated photonic processors}},
    year = {2024},
    journal = {Nature Photonics},
    author = {SeyedinNavadeh, SeyedMohammad and Milanizadeh, Maziyar and Zanetto, Francesco and Ferrari, Giorgio and Sampietro, Marco and Sorel, Marc and Miller, David A B and Melloni, Andrea and Morichetti, Francesco},
    number = {2},
    pages = {149--155},
    volume = {18},
    url = {https://doi.org/10.1038/s41566-023-01330-w},
    doi = {10.1038/s41566-023-01330-w},
    issn = {1749-4893}
}

@article{Russell2017,
    title = {{Direct dialling of Haar random unitary matrices}},
    year = {2017},
    journal = {New Journal of Physics},
    author = {Russell, Nicholas J and Chakhmakhchyan, Levon and O'Brien, Jeremy L and Laing, Anthony},
    number = {3},
    month = {3},
    pages = {33007},
    volume = {19},
    publisher = {IOP Publishing},
    url = {https://iopscience.iop.org/article/10.1088/1367-2630/aa60ed https://iopscience.iop.org/article/10.1088/1367-2630/aa60ed/meta},
    doi = {10.1088/1367-2630/AA60ED},
    issn = {1367-2630},
    arxivId = {1506.06220}
}

@article{Reck1994,
    title = {{Experimental realization of any discrete unitary operator}},
    year = {1994},
    journal = {Physical Review Letters},
    author = {Reck, Michael and Zeilinger, Anton and Bernstein, Herbert J and Bertani, Philip},
    number = {1},
    month = {7},
    pages = {58--61},
    volume = {73},
    publisher = {American Physical Society},
    url = {https://link.aps.org/doi/10.1103/PhysRevLett.73.58},
    doi = {10.1103/PhysRevLett.73.58}
}

@article{Perez-Lopez2024,
    title = {{General-purpose programmable photonic processor for advanced radiofrequency applications}},
    year = {2024},
    journal = {Nature Communications 2024 15:1},
    author = {P{\'{e}}rez-L{\'{o}}pez, Daniel and Gutierrez, Ana and S{\'{a}}nchez, David and L{\'{o}}pez-Hern{\'{a}}ndez, Aitor and Gutierrez, Mikel and S{\'{a}}nchez-Gom{\'{a}}riz, Erica and Fern{\'{a}}ndez, Juan and Cruz, Alejandro and Quir{\'{o}}s, Alberto and Xie, Zhenyun and Benitez, Jesús and Bekesi, Nandor and Santom{\'{e}}, Alejandro and P{\'{e}}rez-Galacho, Diego and DasMahapatra, Prometheus and Macho, Andrés and Capmany, José},
    number = {1},
    month = {2},
    pages = {1--11},
    volume = {15},
    publisher = {Nature Publishing Group},
    url = {https://www.nature.com/articles/s41467-024-45888-7},
    doi = {10.1038/s41467-024-45888-7},
    issn = {2041-1723},
    pmid = {38378716}
}

@article{PhysRevApplied.22.054011,
    title = {{Global calibration of large-scale photonic integrated circuits}},
    year = {2024},
    journal = {Phys. Rev. Appl.},
    author = {Zheng, Jin-Hao and Wang, Qin-Qin and Feng, Lan-Tian and Ding, Yu-Yang and Xu, Xiao-Ye and Ren, Xi-Feng and Li, Chuan-Feng and Guo, Guang-Can},
    number = {5},
    month = {11},
    pages = {54011},
    volume = {22},
    publisher = {American Physical Society},
    url = {https://link.aps.org/doi/10.1103/PhysRevApplied.22.054011},
    doi = {10.1103/PhysRevApplied.22.054011}
}

@article{Bandyopadhyay2021,
    title = {{Hardware error correction for programmable photonics}},
    year = {2021},
    journal = {Optica},
    author = {Bandyopadhyay, Saumil and Hamerly, Ryan and Englund, Dirk},
    number = {10},
    pages = {1247--1255},
    volume = {8},
    publisher = {OSA},
    url = {http://opg.optica.org/optica/abstract.cfm?URI=optica-8-10-1247},
    doi = {10.1364/OPTICA.424052}
}

@article{PhysRevA.92.032322,
    title = {{High-fidelity quantum state evolution in imperfect photonic integrated circuits}},
    year = {2015},
    journal = {Phys. Rev. A},
    author = {Mower, Jacob and Harris, Nicholas C and Steinbrecher, Gregory R and Lahini, Yoav and Englund, Dirk},
    number = {3},
    month = {9},
    pages = {32322},
    volume = {92},
    publisher = {American Physical Society},
    url = {https://link.aps.org/doi/10.1103/PhysRevA.92.032322},
    doi = {10.1103/PhysRevA.92.032322}
}

@article{10092945,
    title = {{Integrated Microwave Photonics Coherent Processor for Massive-MIMO Systems in Wireless Communications}},
    year = {2023},
    journal = {IEEE Journal of Selected Topics in Quantum Electronics},
    author = {Romero, Pablo Martínez-Carrasco and Rausell-Campo, José Roberto and P{\'{e}}rez-Galacho, Diego and Li, Xu and Qing, Ting and Wang, Tiangxiang and P{\'{e}}rez-L{\'{o}}pez, Daniel},
    number = {6: Photonic Signal Processing},
    pages = {1--12},
    volume = {29},
    doi = {10.1109/JSTQE.2023.3264434}
}

@article{https://doi.org/10.1002/advs.202401951,
    title = {{Inverse Design of Photonic Surfaces via High throughput Femtosecond Laser Processing and Tandem Neural Networks}},
    year = {2024},
    journal = {Advanced Science},
    author = {Park, Minok and Grb{\v{c}}i{\'{c}}, Luka and Motameni, Parham and Song, Spencer and Singh, Alok and Malagrino, Dante and Elzouka, Mahmoud and Vahabi, Puya H and Todeschini, Alberto and de Jong, Wibe Albert and Prasher, Ravi and Zorba, Vassilia and Lubner, Sean D},
    number = {26},
    pages = {2401951},
    volume = {11},
    url = {https://advanced.onlinelibrary.wiley.com/doi/abs/10.1002/advs.202401951},
    doi = {https://doi.org/10.1002/advs.202401951}
}

@article{Harris:18,
    title = {{Linear programmable nanophotonic processors}},
    year = {2018},
    journal = {Optica},
    author = {Harris, Nicholas C and Carolan, Jacques and Bunandar, Darius and Prabhu, Mihika and Hochberg, Michael and Baehr-Jones, Tom and Fanto, Michael L and Smith, A Matthew and Tison, Christopher C and Alsing, Paul M and Englund, Dirk},
    number = {12},
    pages = {1623--1631},
    volume = {5},
    publisher = {Optica Publishing Group},
    url = {https://opg.optica.org/optica/abstract.cfm?URI=optica-5-12-1623},
    doi = {10.1364/OPTICA.5.001623}
}

@article{PhysRevApplied.11.064044,
    title = {{Matrix Optimization on Universal Unitary Photonic Devices}},
    year = {2019},
    journal = {Phys. Rev. Appl.},
    author = {Pai, Sunil and Bartlett, Ben and Solgaard, Olav and Miller, David A B},
    number = {6},
    month = {6},
    pages = {64044},
    volume = {11},
    publisher = {American Physical Society},
    url = {https://link.aps.org/doi/10.1103/PhysRevApplied.11.064044},
    doi = {10.1103/PhysRevApplied.11.064044}
}

@inproceedings{Kumar2021MitigatingLO,
    title = {{Mitigating linear optics imperfections via port allocation and compilation}},
    year = {2021},
    author = {Kumar, S Praveen and Neuhaus, Leonhard and Helt, Lukas G and Qi, Haoyu and Morrison, Blair and Mahler, Dylan H and Dhand, Ish}
}

@article{Yuan2024MultiheadedTN,
    title = {{Multi-headed tandem neural network approach for non-uniqueness in inverse design of layered photonic structures}},
    year = {2024},
    journal = {Optics {\&} Laser Technology},
    author = {Yuan, Xiaogen and Wang, Shuqin and Gu, Leilei and Xie, Shusheng and Ma, Qiongxiong and Guo, Jianping},
    url = {https://api.semanticscholar.org/CorpusID:269143144}
}

@article{Perez-Lopez2020,
    title = {{Multipurpose self-configuration of programmable photonic circuits}},
    year = {2020},
    journal = {Nature Communications},
    author = {P{\'{e}}rez-L{\'{o}}pez, Daniel and L{\'{o}}pez, Aitor and DasMahapatra, Prometheus and Capmany, José},
    number = {1},
    pages = {6359},
    volume = {11},
    url = {https://doi.org/10.1038/s41467-020-19608-w},
    doi = {10.1038/s41467-020-19608-w},
    issn = {2041-1723}
}

@misc{lopezmarch2026noiseanalogprogrammablephotoniccomputation,
    title = {{Noise in analog programmable-photonic computation}},
    year = {2026},
    author = {L{\'{o}}pez-March, Raúl and Macho-Ortiz, Andrés and Fraile-Pel{\'{a}}ez, Francisco Javier and Capmany, José},
    url = {https://arxiv.org/abs/2604.24541},
    arxivId = {physics.optics/2604.24541}
}

@article{Clements2016,
    title = {{Optimal design for universal multiport interferometers}},
    year = {2016},
    journal = {Optica},
    author = {Clements, William R and Humphreys, Peter C and Metcalf, Benjamin J and Kolthammer, W Steven and Walmsley, Ian A},
    number = {12},
    pages = {1460--1465},
    volume = {3},
    publisher = {OSA},
    url = {http://www.osapublishing.org/optica/abstract.cfm?URI=optica-3-12-1460},
    doi = {10.1364/OPTICA.3.001460}
}

@misc{akiba2019optunanextgenerationhyperparameteroptimization,
    title = {{Optuna: A Next-generation Hyperparameter Optimization Framework}},
    year = {2019},
    author = {Akiba, Takuya and Sano, Shotaro and Yanase, Toshihiko and Ohta, Takeru and Koyama, Masanori},
    url = {https://arxiv.org/abs/1907.10902},
    arxivId = {cs.LG/1907.10902}
}

@article{Miller:15,
    title = {{Perfect optics with imperfect components}},
    year = {2015},
    journal = {Optica},
    author = {Miller, David A B},
    number = {8},
    pages = {747--750},
    volume = {2},
    publisher = {Optica Publishing Group},
    url = {https://opg.optica.org/optica/abstract.cfm?URI=optica-2-8-747},
    doi = {10.1364/OPTICA.2.000747}
}

@inproceedings{Cavicchioli:24,
    title = {{Programmable integrated photonic circuit for matrix inversion}},
    year = {2024},
    booktitle = {Optical Fiber Communication Conference (OFC) 2024},
    author = {Cavicchioli, G and Miller, D A B and Engheta, N and Melloni, A and Morichetti, F},
    pages = {Th1A.2},
    publisher = {Optica Publishing Group},
    url = {https://opg.optica.org/abstract.cfm?URI=OFC-2024-Th1A.2},
    doi = {10.1364/OFC.2024.Th1A.2}
}

@article{Bogaerts2020,
    title = {{Programmable photonic circuits}},
    year = {2020},
    journal = {Nature},
    author = {Bogaerts, Wim and P{\'{e}}rez, Daniel and Capmany, José and Miller, David A B and Poon, Joyce and Englund, Dirk and Morichetti, Francesco and Melloni, Andrea},
    number = {7828},
    pages = {207--216},
    volume = {586},
    url = {https://doi.org/10.1038/s41586-020-2764-0},
    doi = {10.1038/s41586-020-2764-0},
    issn = {1476-4687}
}

@article{10.1063/5.0235712,
    title = {{Programming universal unitary transformations on a general-purpose silicon photonic platform}},
    year = {2025},
    journal = {APL Photonics},
    author = {Rausell-Campo, José Roberto and P{\'{e}}rez-L{\'{o}}pez, Daniel and Capmany Francoy, José},
    number = {2},
    pages = {26102},
    volume = {10},
    url = {https://doi.org/10.1063/5.0235712},
    doi = {10.1063/5.0235712},
    issn = {2378-0967}
}

@article{Harris2017,
    title = {{Quantum transport simulations in a programmable nanophotonic processor}},
    year = {2017},
    journal = {Nature Photonics},
    author = {Harris, Nicholas C and Steinbrecher, Gregory R and Prabhu, Mihika and Lahini, Yoav and Mower, Jacob and Bunandar, Darius and Chen, Changchen and Wong, Franco N C and Baehr-Jones, Tom and Hochberg, Michael and Lloyd, Seth and Englund, Dirk},
    number = {7},
    pages = {447--452},
    volume = {11},
    url = {https://doi.org/10.1038/nphoton.2017.95},
    doi = {10.1038/nphoton.2017.95},
    issn = {1749-4893}
}

@article{Fan:25,
    title = {{Rapid configuring method for a programmable photonic integrated circuit based on a tandem neural network}},
    year = {2025},
    journal = {Opt. Lett.},
    author = {Fan, Zeyang and Dan, Yihang and Lin, Junmin and Zhang, Tian and Dai, Jian and Xu, Kun},
    number = {5},
    month = {3},
    pages = {1731--1734},
    volume = {50},
    publisher = {Optica Publishing Group},
    url = {https://opg.optica.org/ol/abstract.cfm?URI=ol-50-5-1731},
    doi = {10.1364/OL.551119}
}

@article{Fyrillas:24,
    title = {{Scalable machine learning-assisted clear-box characterization for optimally controlled photonic circuits}},
    year = {2024},
    journal = {Optica},
    author = {Fyrillas, Andreas and Faure, Olivier and Maring, Nicolas and Senellart, Jean and Belabas, Nadia},
    number = {3},
    month = {3},
    pages = {427--436},
    volume = {11},
    publisher = {Optica Publishing Group},
    url = {https://opg.optica.org/optica/abstract.cfm?URI=optica-11-3-427},
    doi = {10.1364/OPTICA.512148}
}

@article{Miller:13,
    title = {{Self-configuring universal linear optical component}},
    year = {2013},
    journal = {Photon. Res.},
    author = {Miller, David A B},
    number = {1},
    month = {6},
    pages = {1--15},
    volume = {1},
    publisher = {Optica Publishing Group},
    url = {https://opg.optica.org/prj/abstract.cfm?URI=prj-1-1-1},
    doi = {10.1364/PRJ.1.000001}
}

@article{Luo:24,
    title = {{Tandem neural network-assisted inverse design of highly efficient diffractive slanted waveguide grating}},
    year = {2024},
    journal = {Opt. Express},
    author = {Luo, Menglong and Lee, Sang-Shin},
    number = {7},
    month = {3},
    pages = {12587--12600},
    volume = {32},
    publisher = {Optica Publishing Group},
    url = {https://opg.optica.org/oe/abstract.cfm?URI=oe-32-7-12587},
    doi = {10.1364/OE.514502}
}

@article{Wu:25,
    title = {{Tandem neural networks for designing nanoplasmonic structures to manipulate quantum dynamics}},
    year = {2025},
    journal = {Opt. Express},
    author = {Wu, Jie and Liu, Guangxin and Li, Lingyan and Xu, Haitao and Lu, Yuwei and Tan, Jie and Liu, Jingfeng},
    number = {5},
    month = {3},
    pages = {9790--9803},
    volume = {33},
    publisher = {Optica Publishing Group},
    url = {https://opg.optica.org/oe/abstract.cfm?URI=oe-33-5-9790},
    doi = {10.1364/OE.553233}
}

@inproceedings{10416579,
    title = {{Tandem Neural Networks for the Inverse Programming of Linear Photonic Processors}},
    year = {2023},
    booktitle = {2023 International Topical Meeting on Microwave Photonics (MWP)},
    author = {Rausell-Campo, José Roberto and P{\'{e}}rez-L{\'{o}}pez, Daniel and Shastri, Bhavin and Melati, Daniele},
    pages = {1--4},
    doi = {10.1109/MWP58203.2023.10416579}
}

@article{Wang:20,
    title = {{Tolerant, broadband tunable 2x2 coupler circuit}},
    year = {2020},
    journal = {Opt. Express},
    author = {Wang, Mi and Ribero, Antonio and Xing, Yufei and Bogaerts, Wim},
    number = {4},
    month = {2},
    pages = {5555--5566},
    volume = {28},
    publisher = {Optica Publishing Group},
    url = {https://opg.optica.org/oe/abstract.cfm?URI=oe-28-4-5555},
    doi = {10.1364/OE.384018}
}

@article{doi:10.1126/science.aab3642,
    title = {{Universal linear optics}},
    year = {2015},
    journal = {Science},
    author = {Carolan, Jacques and Harrold, Christopher and Sparrow, Chris and Mart{\'{i}}n-L{\'{o}}pez, Enrique and Russell, Nicholas J and Silverstone, Joshua W and Shadbolt, Peter J and Matsuda, Nobuyuki and Oguma, Manabu and Itoh, Mikitaka and Marshall, Graham D and Thompson, Mark G and Matthews, Jonathan C F and Hashimoto, Toshikazu and O’Brien, Jeremy L and Laing, Anthony},
    number = {6249},
    pages = {711--716},
    volume = {349},
    url = {https://www.science.org/doi/abs/10.1126/science.aab3642},
    doi = {10.1126/science.aab3642}
}

@article{Annoni2017,
    title = {{Unscrambling light—automatically undoing strong mixing between modes}},
    year = {2017},
    journal = {Light: Science {\&} Applications},
    author = {Annoni, Andrea and Guglielmi, Emanuele and Carminati, Marco and Ferrari, Giorgio and Sampietro, Marco and Miller, David A B and Melloni, Andrea and Morichetti, Francesco},
    number = {12},
    pages = {e17110-e17110},
    volume = {6},
    url = {https://doi.org/10.1038/lsa.2017.110},
    doi = {10.1038/lsa.2017.110},
    issn = {2047-7538}
}

@article{Zhou2025AComputing,
    title = {{A manufacturable platform for photonic quantum computing}},
    year = {2025},
    journal = {Nature 2025 641:8064},
    author = {Zhou, Xinran and Wooding, Jamie and Wingert, Matthew and Weigel, Peter and Vorobeichik, Ilya and Vidrighin, Mihai D. and Vert, Alexey and Veitia, Andrzej and Tung, Maryann and Triplett, Mark and Tran, Khanh and Thompson, Mark G. and Tamborini, Davide and Sukumaran, Vijay and Stavrakas, Camille and Staffaroni, Matteo and Sparrow, Chris and Souza, Mario C.M.M. and Son, Gyeongho and Sohn, Young Ik and Sohn, Ben and Smith, Jake and Sinsky, Jeffrey and Shin, Hyungki and Bahgat Shehata, Andrea and Shah, Deesha and Shadbolt, Pete and Saunders, Dylan J. and Rudolph, Terry and Roxworthy, Brian and Rodriguez, Angelita Viejo and Ray, Gareth and Ramprasad, Tarun and Pryde, Geoff J. and Poush, Matt and Peterson, Gabriel and Peterson, Brennan and Penthorn, Nicholas and Peng, Hsuan Tung and Park, Bryan and Pai, Sunil and Ortmann, J. Elliott and O’Brien, Jeremy L. and Najafi, Faraz and Musalem, Francois Xavier and Munns, Joseph and Mukherjee, Shaunak and Moores, Brad and Mendoza, Gabriel and Melnichuk, Ann and Manfrinato, Vitor Riseti and Lovelady, Michael and Lokovic, Kimberly and Llewellyn, Dan and LiCausi, Nicholas and Liang, Yong and Kumar, Nikhil and Krogen, Peter and Kovall, George and Koustuban, Ravi and Kelez, Nicholas and Kamineni, Vimal and Jones, Thomas and Johansson, Henrik and Jain, Vijay and Jadidi, Mehdi and Hu, Hong and House, Matthew and Horng, Jason and Hardy, Sam and Hansen, Paul and Halimi, Sami and Haislmaier, Ryan and Goley, Patrick and Goeldi, Sebastian and Gimeno-Segovia, Mercedes and Gibson, Gary and Ganesan, Yogeeswaran and Fukami, Masaya and Frazer, Jonathan and Fenrich, Colleen and Farsi, Alessandro and Fargas, Josep and Er-Xuan, Ping and Dudley, Eric and Davis, Michael and Dauer, Tom and Danesh, Fariba and Chung, C. J. and Choudhury, Sourav Sen and Chang, Chia Ming and Ceballos, Alejandro and Catalano, Gabriel and Campbell, Geoff and Cable, Hugo and Burridge, Ben and Burgos, Stanley and Bonneau, Damien and Black, Dylan and Benyamini, Avishai and Alexander, Koen},
    number = {8064},
    month = {2},
    pages = {876--883},
    volume = {641},
    publisher = {Nature Publishing Group},
    url = {https://www.nature.com/articles/s41586-025-08820-7},
    doi = {10.1038/s41586-025-08820-7},
    issn = {1476-4687},
    pmid = {40010377},
    arxivId = {2404.17570}
}

@article{Zibar2023AddressingLearning,
    title = {{Addressing data scarcity in optical matrix multiplier modeling using transfer learning}},
    year = {2023},
    journal = {Optics Letters, Vol. 48, Issue 24, pp. 6553-6556},
    author = {Zibar, Darko and Jovanovic, Ognjen and Yan, Siqi and Cem, Ali and Ding, Yunhong and Ros, Francesco Da},
    number = {24},
    month = {12},
    pages = {6553--6556},
    volume = {48},
    publisher = {Optica Publishing Group},
    url = {https://opg.optica.org/viewmedia.cfm?uri=ol-48-24-6553&seq=0&html=true https://opg.optica.org/abstract.cfm?uri=ol-48-24-6553 https://opg.optica.org/ol/abstract.cfm?uri=ol-48-24-6553},
    doi = {10.1364/ol.502517},
    issn = {0146-9592},
    pmid = {38099797}
}

@article{Macho-Ortiz2026AnalogInformation,
    title = {{Analog programmable-photonic information}},
    year = {2026},
    journal = {Advanced Photonics},
    author = {Macho-Ortiz, Andrés and L{\'{o}}pez-March, Raúl and Romero, Pablo Martínez-Carrasco and Fraile-Pel{\'{a}}ez, Francisco Javier and Capmany, José},
    number = {3},
    month = {4},
    pages = {36002},
    volume = {8},
    url = {https://doi.org/10.1117/1.AP.8.3.036002},
    doi = {10.1117/1.AP.8.3.036002}
}

@article{Hamerly2022AsymptoticallyPhotonics,
    title = {{Asymptotically fault-tolerant programmable photonics}},
    year = {2022},
    journal = {Nature Communications},
    author = {Hamerly, Ryan and Bandyopadhyay, Saumil and Englund, Dirk},
    number = {1},
    pages = {6831},
    volume = {13},
    url = {https://doi.org/10.1038/s41467-022-34308-3},
    doi = {10.1038/s41467-022-34308-3},
    issn = {2041-1723}
}

@article{Kim2023DeepArrays,
    title = {{Deep neural network-based phase calibration in integrated optical phased arrays}},
    year = {2023},
    journal = {Scientific Reports},
    author = {Kim, Jae-Yong and Kim, Junhyeong and Yoon, Jinhyeong and Hong, Seokjin and Neseli, Berkay and Kwon, Namhyun and You, Jong-Bum and Yoon, Hyeonho and Park, Hyo-Hoon and Kurt, Hamza},
    number = {1},
    pages = {19929},
    volume = {13},
    url = {https://doi.org/10.1038/s41598-023-47004-z},
    doi = {10.1038/s41598-023-47004-z},
    issn = {2045-2322}
}

@article{Youssry2024ExperimentalControl,
    title = {{Experimental graybox quantum system identification and control}},
    year = {2024},
    journal = {npj Quantum Information},
    author = {Youssry, Akram and Yang, Yang and Chapman, Robert J and Haylock, Ben and Lenzini, Francesco and Lobino, Mirko and Peruzzo, Alberto},
    number = {1},
    pages = {9},
    volume = {10},
    url = {https://doi.org/10.1038/s41534-023-00795-5},
    doi = {10.1038/s41534-023-00795-5},
    issn = {2056-6387}
}

@article{Milanizadeh2022SeparatingProcessor,
    title = {{Separating arbitrary free-space beams with an integrated photonic processor}},
    year = {2022},
    journal = {Light: Science {\&} Applications},
    author = {Milanizadeh, Maziyar and SeyedinNavadeh, SeyedMohammad and Zanetto, Francesco and Grimaldi, Vittorio and De Vita, Christian and Klitis, Charalambos and Sorel, Marc and Ferrari, Giorgio and Miller, David A B and Melloni, Andrea and Morichetti, Francesco},
    number = {1},
    pages = {197},
    volume = {11},
    url = {https://doi.org/10.1038/s41377-022-00884-8},
    doi = {10.1038/s41377-022-00884-8},
    issn = {2047-7538}
}

@article{Bandyopadhyay2024Single-chipTraining,
    title = {{Single-chip photonic deep neural network with forward-only training}},
    year = {2024},
    journal = {Nature Photonics},
    author = {Bandyopadhyay, Saumil and Sludds, Alexander and Krastanov, Stefan and Hamerly, Ryan and Harris, Nicholas and Bunandar, Darius and Streshinsky, Matthew and Hochberg, Michael and Englund, Dirk},
    number = {12},
    pages = {1335--1343},
    volume = {18},
    url = {https://doi.org/10.1038/s41566-024-01567-z},
    doi = {10.1038/s41566-024-01567-z},
    issn = {1749-4893}
}


\end{document}